\documentclass[twocolumn,twoside,slac_two]{revtex4}
\pdfoutput=0
\usepackage{graphicx}
\usepackage{fancyhdr}
\usepackage{graphics}
\usepackage{epstopdf}
\usepackage{textpos}
\usepackage{textcomp}
\usepackage{natbib}
\begin{document}

%%%%%%%%%%%%%%%%%%%%%% WRITE THE TITLE HERE %%%%%%%%%%%%%%%%%%%
\title{\centering W/Z and diboson production at hadron colliders}
%%%%%%%%%%%%%%%%%%%%%% WRITE THE AUTHOR HERE %%%%%%%%%%%%%%%%%
\author{
\centering
\includegraphics[scale=0.2]{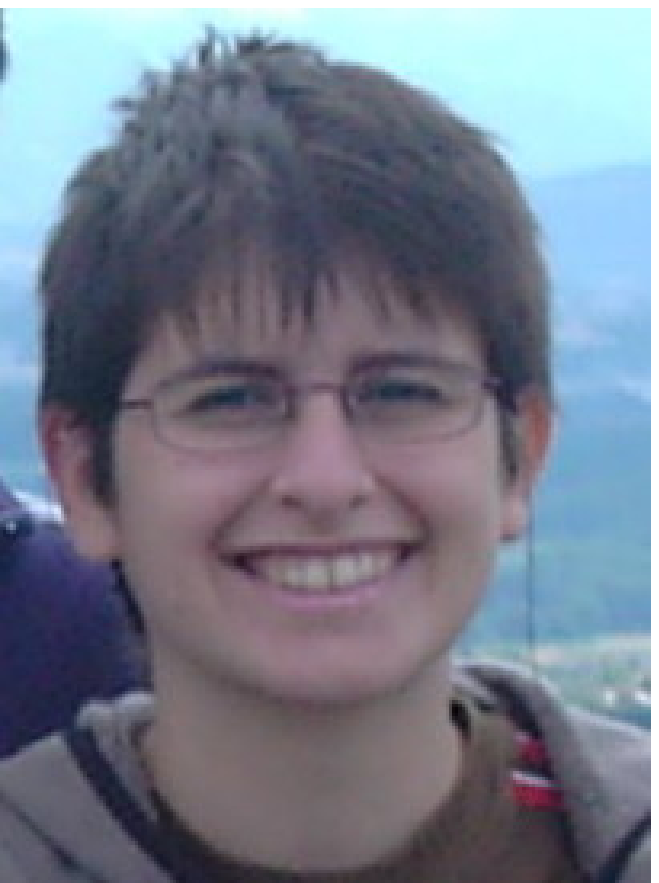} \\
\begin{center}
Sara Bolognesi on behalf of ATLAS, CDF, CMS and D0 collaborations
\end{center}}
\affiliation{\centering Johns Hopkins University and CERN}
%%%%%%%%%%%%%%%%%%%%%% WRITE THE ABSTRACT HERE %%%%%%%%%%%%%%%%
\begin{abstract}
A general review of the latest results about single and double vector boson production
in the multipurpose experiments at LHC~\cite{Evans:2008zzb} (ATLAS~\cite{Aad:2008zzm} and CMS~\cite{CMSDetector}) 
and at Tevatron~\cite{Wilson:1977nk} (CDF~\cite{Blair:1996kx} and D0~\cite{Abazov:2005pn}) will be presented.
The review will focus on boson production, while a more detailed report
about $W$ and $Z$ properties can be found elsewhere~\cite{Melnitchouk:2011tq}. 
Only leptonic decays into electrons and muons will be considered.
\end{abstract}

\maketitle
\thispagestyle{fancy}


\section{Introduction}
In the complex environment of hadron colliders, 
$W$ and $Z$ events with leptonic decays have a clear signature and the 
processes of boson production and decay are relatively well known.
For these reasons, they are heavily exploited for detector commissioning 
in the early stage of data taking.
$W$ and $Z$ are also the main handles to perform precision
Standard Model (SM) measurements 
which are important as input to the Global Electroweak fit~\cite{Baak:2011ze} and to
constraint the Parton Distribution Functions~\cite{Chlebana:2005sd,Watt:2011kp}.
At the present hadron colliders the rate of single boson production is huge:
about 10 (3) Millions of $W\rightarrow l\nu$ are produced per $fb^{-1}$
in each experiment at LHC (Tevatron), the rate for $Z \rightarrow ll$ events being about $1/10$ of that.
The main challenge is the control of systematics in presence
of pile-up events and given an initial state not completely known.

Single boson production in association with jets and diboson production
are the main backgrounds for most of the Higgs channel and generic New Physics searches.
It is of fundamental importance to control these backgrounds with very good precision.
They also play a crucial role as benchmark test to establish the analysis techniques
to search for Higgs and/or New Physics.

After a brief review, in Section~\ref{singleboson}, about the bosons signature and recent inclusive 
cross-section measurements, Section~\ref{inassociation} will focus on measurements sensitive to 
higher perturbative $\alpha_S$ orders (NLO, NNLO) in the boson production process, like 
$Z$ transverse momentum ($p_T$) distribution and $W/Z$ production in association with jets
(first measurements in association with heavy-flavor jets will also be reported).
The production of a boson in association with a photon will be reviewed in Section~\ref{withphoton}.
Finally, latest measurements of diboson production ($WW$,$WZ$,$ZZ$) will be reported in Section~\ref{diboson}.
%%ANYTHING ABOUT CONCLUSIONS???

\section{$W$ and $Z$ signatures and inclusive cross sections}
\label{singleboson}
$W$ ($Z$) events are characterized by the presence of one (two) isolated lepton
with relatively high $p_T$, which is exploited to trigger the event on-line.
In case of $W$ events, the undetected neutrino gives rise to an unbalancing in the total
transverse energy ($E_T^{miss}$). On top of this typical features, additional hadronic activity
due to pile-up and underlying event, as well as hard jets due to NLO contributions,
are present. The $W$ signal is usually extracted from a fit to the $E_T^{miss}$ distribution
(or the transverse mass distribution built with $E_T^{miss}$ and lepton $p_T$), while $Z$ signal
is easily identifiable thanks to the invariant mass peak of the two leptons.
The main background is due to QCD processes with one (or two) real leptons from
heavy-flavor decays and/or fake leptons. The QCD background is rejected by requiring the
leptons to be well identified and isolated (i.e., far away from any hadronic activity).
The remaining QCD contamination is measured from control regions in data (e.g., an anti-isolated sample),
since its simulation with simple LO and Parton Shower (PS) Monte Carlo (MC) is not reliable and
the lepton fake rate is very much dependent on the details of the detector.
The remaining background for $W$ is due to Electroweak processes (e.g., $Z$, double top, diboson, 
tau decays of $W$), where part of the event is not reconstructed correctly
or it falls outside the detector acceptance. These processes are relatively
well known and their contribution is extracted from simulation.

An inclusive $W$ and $Z$ signal extraction has been performed on the
first data by CMS~\cite{WZCMS:2011nx} and ATLAS~\cite{Aad:2011dm}.
The measured cross sections are in good agreement with theoretical prediction at 
NNLO based on recent parton distribution functions.
With 35~$pb^{-1}$ the statistical error is already negligible (few \textperthousand).
With the exception of the luminosity uncertainty (4\%), the other
experimental systematic uncertainties (which are dominated by the $E_T^{miss}$ resolution
and the lepton efficiencies) are lower ($\simeq$1\%) than the theoretical uncertainty
($\leq$3\%). Part of the theoretical error is due to higher perturbative orders
(computations are available up to QCD NNLO and EWK NLO) and to Initial and
Final State Radiation (ISR and FSR). This theoretical uncertainty can be further investigated
with differential cross section measurements, as will be shown in Section~\ref{inassociation}.
The other contribution to the theoretical uncertainty is due to PDF.
On the other hand, the argument can be reversed and the cross section measurement
inside the fiducial region can be exploited to constrain further the PDF.
Useful measurements, in this regard, are the ratio between $W^+$ and $W^-$ cross sections
and the rapidity distribution of the bosons~\cite{Chatrchyan:2011wt,Aad:2011dm,Han:2011vw}. More details about
PDFs and the latest related measurements from LHC and Tevatron can be found
in~\cite{Placakyte:2011az}.

\section{Beyond Leading Order: transverse momentum distribution and production in association with jets}
\label{inassociation}
\begin{figure} [ht!]
\includegraphics[width=80mm]{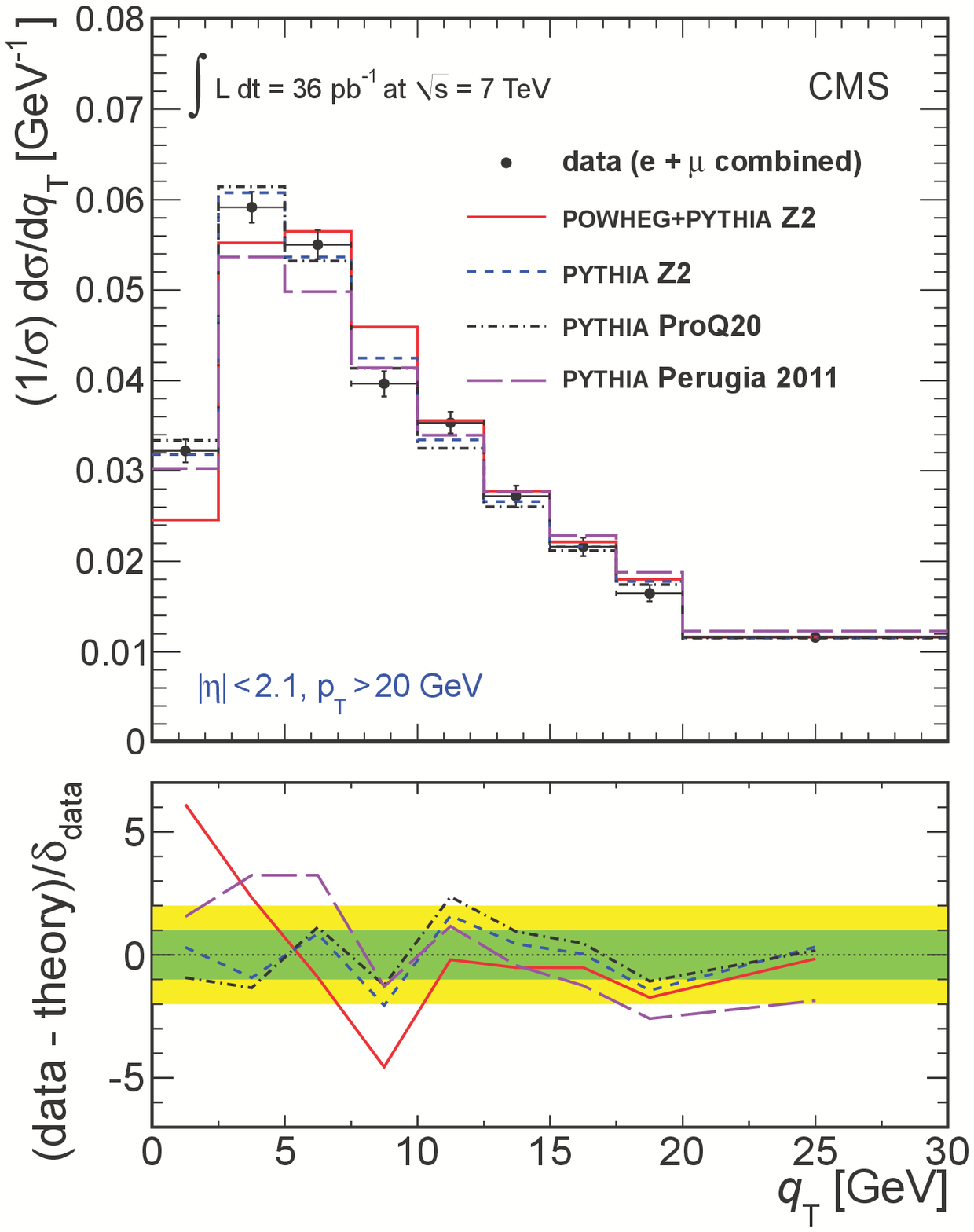}
\includegraphics{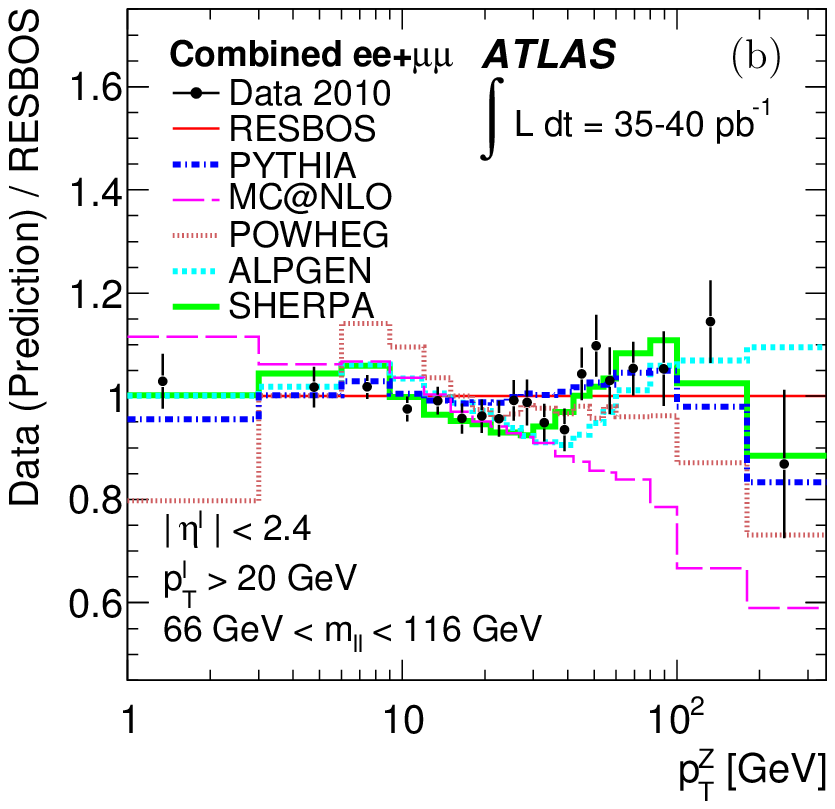}
\caption{$Z$ transverse momentum distribution. Top: low transverse momentum region measured
by CMS compared to several Pythia tunes (the green and yellow bands are $1\sigma$ and $2\sigma$
experimental uncertainty). Bottom: ATLAS $Z$ transverse momentum measurement compared with various theoretical predictions 
(data points are shown with combined statistical and systematic uncertainty).}
\label{Zpt}
\end{figure}

At LO the boson is produced with null $p_T$ while at higher perturbative orders
the boson gets boosted recoiling against additional produced jets.
More precisely, two $p_T$ regions may be identified: the low $p_T$ region, which
is dominated by non-perturbative effects and multiple soft gluon radiation and the
high $p_T$ region, dominated by hard gluon emission well described by perturbative QCD.
ATLAS and CMS have measured the $Z$ $p_T$~\cite{Chatrchyan:2011wt,Aad:2011gj} with
35 $fb^{-1}$, as shown in Fig.~\ref{Zpt}. On the top the sensitivity to the
Underlying Event tune in the low $p_T$ region is shown, on the bottom
the comparison between data and various theoretical prediction is shown.
As can be seen, the NLO predictions (MC@NLO~\cite{Frixione:2002ik}, POWHEG~\cite{Frixione:2007vw}) fail
at high $p_T$ while the NNLO prediction (ResBos~\cite{Balazs:1997xd}) shows good agreement with
data. Including also Next-to-Next-to-Leading-Log resummation, ResBos is
able to describe the data down to very low $p_T$. Another theoretical approach
consists in describing the LO contribution with Matrix Element
at different jet multiplicities and matching this computation to a PS MC for the
description of the soft radiation. This matching procedure is implemented
in Alpgen~\cite{Mangano:2002ea}, Madgraph~\cite{Alwall:2007st} and Sherpa~\cite{Gleisberg:2008ta} which
are able to describe well the full boson $p_T$ spectrum, as well as, the
jet multiplicity.

\begin{figure*}[ht!]
\includegraphics{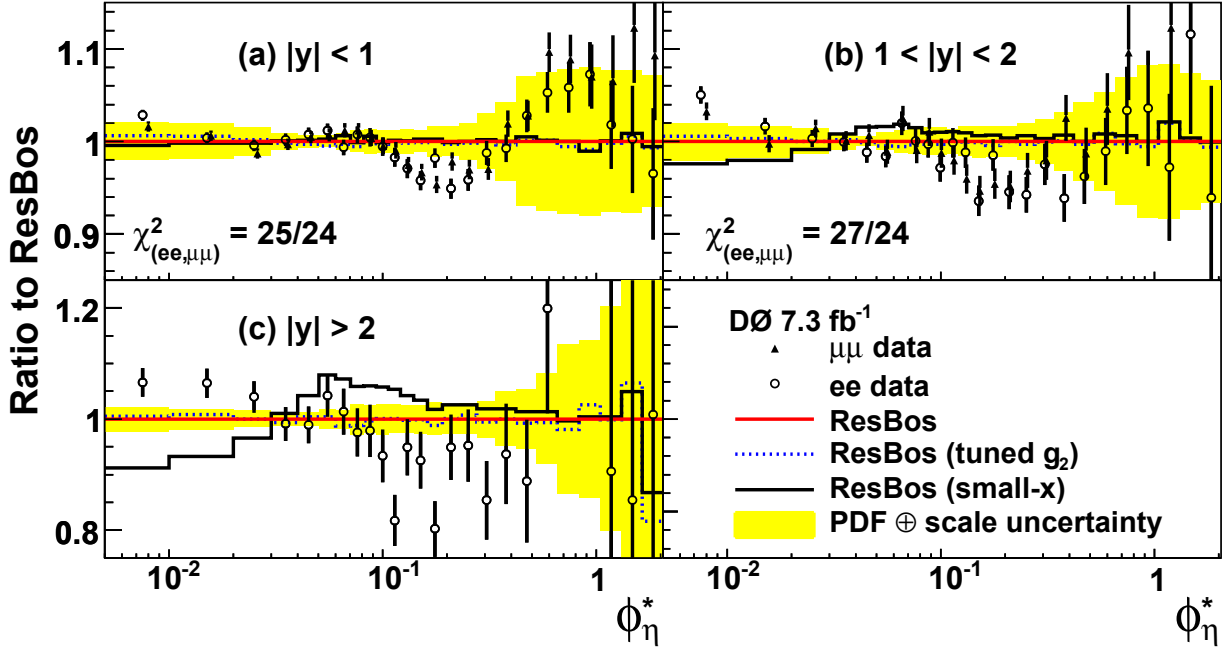}
\caption{Angular variable which is most sensitive to the $Z$ transverse momentum as measured by D0 
in different $Z$ rapidity regions. Statistical and systematic uncertainties are combined in quadrature. 
In the common acceptance region for electrons and muons a $\chi^2$ estimating the agreement 
between the measurements in the two channels is calculated assuming uncorrelated uncertainties. 
Measurements are compared with ResBos prediction ($g_2$ is a parameter of the MC which controls 
a non-perturbative form factor and ``small-x'' refers to an increase of this form
factor for $x < 10^{-2}$, where x is the parton momentum fraction).The yellow band around the
ResBos prediction represents the quadrature sum of uncertainty due to PDFs and the uncertainty 
due to the QCD scale.}
\label{Zangles}
\end{figure*}
Experimentally, the uncertainty on the $p_T$ measurement is dominated by the
lepton scale. To improve the sensitivity, D0 has developed a new
technique which is based on the measurement of pure angular variables~\cite{Abazov:2010mk}:
once identified the $Z$ $p_T$ component most sensitive to the production mechanism, 
this is divided by the dilepton invariant mass to remove the dependence on the energy scale.
This analysis allows to reach the maximum sensitivity with minimal experimental
systematic uncertainty. The results are shown in Fig.~\ref{Zangles}: interesting discrepancies with respect
to ResBos prediction are visible.

\begin{figure}[ht!]
\includegraphics{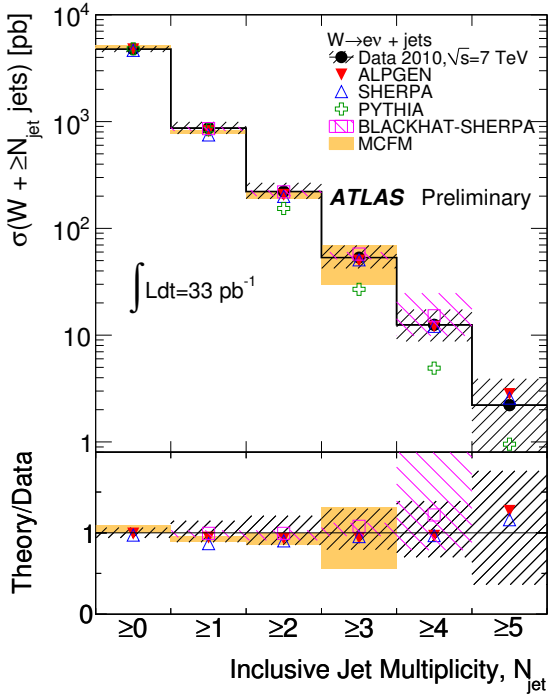}
\caption{$W$+jets cross section results from ATLAS as a function of jet multiplicity in electron channel.
For the data, the statistical uncertainties are shown by the vertical bars and the
combined statistical and systematic uncertainties are shown by the black-hashed regions. Several theoretical 
predictions are compared. The theoretical uncertainties are
shown only for MCFM (NLO prediction for $Njet \leq 2$ and a LO prediction for $Njet = 3$) and BLACKHAT-SHERPA
(NLO prediction for $Njet \leq 3$ and LO prediction for $Njet = 4$).}
\label{ATLASWJets}
\end{figure}

\begin{figure}[ht!]
\includegraphics{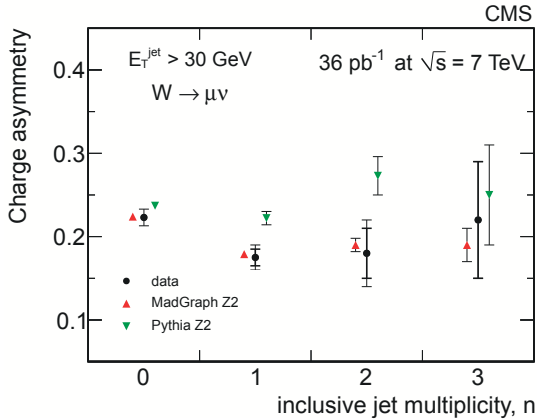}
\caption{$W$ charge asymmetry $\frac{W^+-W^-}{W^++W^-}$ as a function of jet multiplicity
measured by CMS in the muon channel. 
Data are compared to Madgraph and Pythia simulation.
Error bars are shown for the statistical only (black) and total uncertainty (gray).}
\label{CMSWJets}
\end{figure}

Measurement of additional jets recoiling against the boson is, instead, a direct
probe of the higher perturbative orders but at the expense of much higher systematics
uncertainties 
related with jet reconstruction (mainly due to jet energy scale and pile-up removal).
LHC experiments have already acquired a very good control of these uncertainties, as
can be seen in the measurement~\cite{Wjets:2010pg} shown in Fig.~\ref{ATLASWJets}, 
%decidere se mettere anche pt jets
allowing a first successful test of ME-PS matching tools in contrast with the
failing predictions from pure PS MC. The high statistics available
at LHC allowed to reach high jet multiplicity and to perform differential
measurements. On the other hand measurements at low jet $p_T$ are still
limited by systematic uncertainties, 
even at Tevatron~\cite{Abazov:2011rf,ZjetsCDF}. Again, focusing on
angular variables garantees to keep the experimental uncertainty much smaller
than the theoretical uncertainty~\cite{Abazov:2009pp}. Another way to minimize the experimental
systematic uncertainties is to measure ratios between rates~\cite{Chatrchyan:2011ne}, 
as can be seen in Fig.~\ref{CMSWJets}.
This approach has the advantage that also many of the theoretical uncertainties
cancel out, therefore any deviation from the expected behavior is a direct hint of New
Physics.


For similar reasons, the measurement of boson production in association with heavy-flavor
jets is usually reported as ratio with respect to the flavor-inclusive measurement, such that
the only remaining systematic error comes from the heavy-flavor tagging. The estimation of
b-tag efficiency and fake rate are extracted from data and the level of control
of these systematic uncertainties at LHC is already very good, allowing to perform first measurement
of $Z+b$~\cite{Aad:2011jn,ZbjetsCMS}, $W+b$~\cite{Aad:2011kp} and $W+c$~\cite{WcjetsCMS}. 
These measurement are particularly interesting 
to constraint PDF for non-valence quarks (respectively, b and s quarks). Moreover 
they constitute an important background, as well as benchmark analyses, for Higgs searches 
in SM channels like $ZH\rightarrow llbb$ or SuperSymmetric channels like $\phi bb$.
This interest motivates also recent measurements performed at Tevatron~\cite{Abazov:2010ix,Abazov:2008qz,ZbjetsCDF,WcjetsCDF}.

\section{Single vector boson production in association with a photon}
\label{withphoton}
The production of a boson in association with a photon may come through 
ISR or FSR or through Triple Gauge Coupling (TGC), the $ZZ\gamma$ and $Z\gamma\gamma$
vertexes being null at LO in SM. This last contribution is the most interesting one
for a stringent test of SM.

\begin{figure}[h!]
\includegraphics[width=80mm]{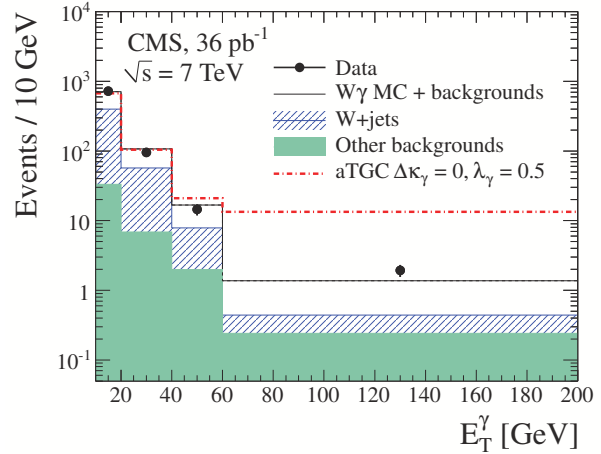}
\caption{Transverse energy distribution for the photon candidates for $W\gamma$ production: CMS data are
compared to MC expectations. A typical aTGC signal is given as a red dot-and-line histogram. The last bin
includes overflows. Entries in wider bins are normalized to the ratio of 10 GeV and the bin
width.}
\label{CMS_Wgamma}
\end{figure}

The photon is reconstructed as an isolated deposit in the electromagnetic calorimeter with
the expected showering shape and no track associated. Usually photons of
low transverse energy ($<10$ GeV) and near to a lepton are rejected to avoid
the divergence of the LO cross section. Moreover, photons are required to be isolated to remove 
the ones coming from hadron fragmentation. Measurement done at LHC~\cite{Chatrchyan:2011rr,Aad:2011tc} 
and Tevatron~\cite{Abazov:2011rk,Aaltonen:2011zc}
are in good agreement with SM expectations, as can be seen, for instance, in Fig.~\ref{CMS_Wgamma}.

\section{Diboson production}
\label{diboson}
The production of a couple of vector bosons ($VV=WW$,$ZZ$,$ZW$) may happen at LO
from $qq$ initial state through exchange of a quark into t-channel or through TGC.
Again, since the TGC is completely fixed by the Electroweak gauge structure of the theory,
any deviation from SM expectations in the VV spectrum is a direct hint of New Physics
in the gauge bosons sector. 
At NLO also the VV production from $gg$ initial state through a quark box is allowed.
The contribution is only a few \% but with large ($\sim$ 50\%) theoretical uncertainty. The precise
knowledge of VV production is very much relevant since this is the main irreducible
background to Higgs production at high mass. The VV analyses are therefore benchmark
tests to establish the Higgs search techniques in this region.

\begin{figure}[h!]
\includegraphics[width=80mm]{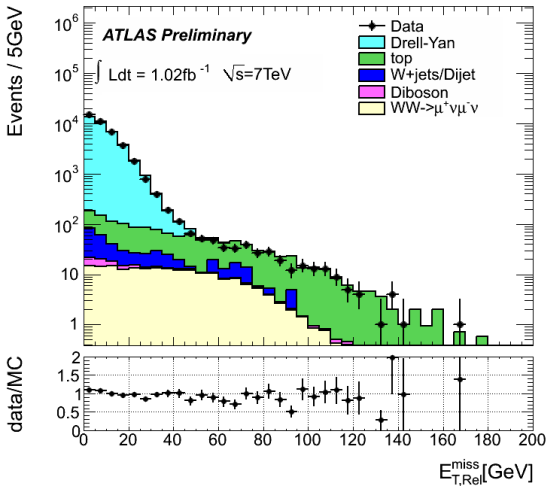}
\caption{Missing transverse energy distributions in ATLAS dimuon events selected for $WW$ analysis.}
\label{ATLAS_WW}
\end{figure}

Considering the fully leptonic final states, the highest production rate is in the 
channel $WW\rightarrow l\nu l\nu$. This is a complex analysis since no mass peak is available
and many different backgrounds need to be subtracted. 
The smallest contributions come from other diboson final states and are estimated from simulation.
The highest backgrounds are $W$+jets, top and Drell-Yan processes which may be rejected, respectively, 
by the tight lepton quality requirements, by the (b-)jet veto and by requiring large $E_T^{miss}$.
The remaining contributions from these processes are estimated from data in control regions.
It is mandatory to have a good control on jet veto and leptonic efficiency and
on $E_T^{miss}$ uncertainty to keep the systematic uncertainties on signal acceptance low.
This objective has been reached at LHC, as can be seen by the very good agreement of data and MC
in Fig.~\ref{ATLAS_WW}, and the measured $WW$ cross section~\cite{Yang:2011db,VVCMS} is in good agreement with NLO expectations. 

Similar results have been obtained at LHC for $WZ\rightarrow l\nu ll$ cross section measurements~\cite{VVCMS,WZATLAS}.
This final state is basically background free and it is particularly interesting since, being charged, is accessible
only at hadron colliders. Recent measurement has been performed by Tevatron with very
sophisticated analysis techniques. While the expected cross-section at
Tevatron is 6 times smaller than LHC, D0 has been able to keep the signal statistics 
at half of the LHC results~\cite{Abazov:2010qn}. Moreover, while at LHC diboson analyses are, for now, mainly based
on simple cut-and-count techniques, CDF, for instance, uses a shape analysis with a fit
to a Neural Network output to extract the $WZ$ cross-section~\cite{WZleptCDF}.
Tevatron experiments have also measured semileptonic final states like $WW/WZ\rightarrow l\nu jj$~\cite{Aaltonen:2009vh} 
and $WZ/ZZ\rightarrow l\nu / \nu\nu bb$~\cite{Abazov:2009jf,Aaltonen:2011uj}  which have a huge background coming from
$W/Z$+jets. Again, to boost the sensitivity CDF and D0 exploit sophisticated analysis
strategies like, ME methods, Boosted Decision Trees and Neural Networks. All the results
are in agreement with NLO expectations.

Finally the final state with the lowest statistics but also the smallest background is $ZZ\rightarrow 4l$.
This channel is the most powerful one for the Higgs search at intermediate and high mass. At this moment
LHC and Tevatron experiments each expect of order 10 $ZZ$ events from electroweak continuum~\cite{VVCMS,Aad:2011xj}.
The results are in agreement with NLO expectations and strongly limited by the
statistical uncertainty ($\sim$ 30\%). The most precise measurement of $ZZ$ cross section~\cite{Abazov:2011td}
comes from the combination of $4l$ and $2l2\nu$ final states done at D0.
CDF has also looked to the semileptonic final state $2l2j$  as well as the $2l2\nu$ final state~\cite{WZCDF} 
where the signal is overwhelmed by the huge $Z$+jets process.%, as can be seen in Fig.~\ref{}.

In Table~\ref{VVxsecs} a list of the recent VV cross section measurement from the different experiments is reported.

\begin{table*}[ht]
\begin{center}
\caption{Summary of the most recent VV cross section measurements}
\begin{tabular}{|l|c|c|}
\hline \textbf{Experiment: channel} & \textbf{Cross section measurement (pb)} & \textbf{NLO expectation (pb)} 
\\
\hline
ATLAS: $WW\rightarrow l\nu l\nu $ & $55.3 \pm 3.3(stat)\pm 6.9(syst)\pm 3.3(lumi)$  & 43 $\pm$ 2 \\
CMS: $WW\rightarrow l\nu l\nu $ & $48.2 \pm 4.0(stat)\pm 6.4(syst)\pm 1.8(lumi)$  & 46 $\pm$ 3 \\
\hline
ATLAS: $WZ\rightarrow l\nu ll $ & $21.1^{+3.1}_{-2.8}(stat)\pm 1.2 (syst) ^{+0.9}_{-0.8}(lumi)$ &  $17.2^{+1.2}_{-0.8}$ \\
CMS: $WZ\rightarrow l\nu ll $ & $17.0 \pm 2.4(stat)\pm 1.1(syst)\pm 1.0(lumi)$ &  $18.75^{+1.1}_{-0.8}$ \\
D0: $WZ\rightarrow l\nu ll $ & $3.90^{+1.09}_{-0.90}$ & $3.25\pm 0.19$ \\
CDF: $WZ\rightarrow l\nu ll $ & $3.9^{+0.6}_{-0.5}(stat)^{+0.6}_{-0.4}(syst)$ & $3.46\pm 0.21$ \\
\hline
CDF: $WZ/WW \rightarrow l\nu jj $ & $16.0 \pm 3.3$ & $16.1 \pm 0.9$ \\
CDF: $WZ/ZZ \rightarrow l\nu/\nu\nu bb $ & $5.8^{+3.6}_{-3.0}$ & 5.1 \\
D0: $WZ/ZZ \rightarrow l\nu/\nu\nu bb $ & $6.9\pm 1.3(stat)\pm 1.8(syst)$ & 4.6 \\
\hline
ATLAS: $ZZ\rightarrow llll $ & $8.4^{+2.7}_{-2.3}(stat)^{+0.4}_{-0.7}(syst)\pm 0.3(lumi)$  & $6.5^{+0.3}_{-0.2}$ \\
CMS: $ZZ\rightarrow llll $ & $3.8^{+1.5}_{-1.2}(stat)\pm 0.2(syst)\pm 0.2(lumi)$  & $6.4^{+0.3}_{-0.2}$ \\
CDF: $ZZ\rightarrow llll $ & $2.8^{+1.2}_{-0.9}(stat)\pm 0.3(syst)$  & $1.4\pm 0.1$ \\
\hline
D0: $ZZ\rightarrow ll\nu\nu / lllll $ &  $1.40^{+0.43}_{-0.37}(stat)\pm 0.14(syst)$ & $1.4\pm 0.1$\\
CDF: $ZZ\rightarrow ll\nu\nu $ & $1.45^{+0.45}_{-0.42}(stat)^{+0.41}_{-0.30}(syst)$ & $1.21^{+0.05}_{-0.04}(scale)^{+0.04}_{-0.03}(PDF)$ \\
\hline
\end{tabular}
\label{VVxsecs}
\end{center}
\end{table*}

\section{Conclusions}
The latest results from Tevatron and LHC for single vector boson production, also in association with
jets or with a photon, and for double vector boson production have been reviewed.

Tevatron experiments have today analyzed almost all the statistics available (usually the analyses rely
on 5-7 $fb^{-1}$, while of the order of 10 $fb^{-1}$ are available), CDF and D0 
have the best possible control of the experimental uncertainties and results are extracted
with the most innovative techniques (angular analyses, ME methods, etc.).

LHC has just released the first set of results: the single boson measurement, even
in association with jets or photon, are based on 35~$pb^{-1}$, while the diboson analyses 
exploit 1 $fb^{-1}$. Today $\sim$5 $fb^{-1}$ are available for analysis and at least twice
of that statistics is expected for next year. The control of systematic uncertainties is already at very
good level and room for improvement is still available.  

The $W$ and $Z$ measurements described in this report  are paving the road to the Higgs or New Physics 
discovery in the VV scattering spectrum.

%\begin{acknowledgments}
%Work supported by ?????????????
%\end{acknowledgments}

\bigskip % extra skip inserted
%% Create the reference section using BibTeX:
%\bibliographystyle{unsrt}
%\bibliographystyle{h-elsevier}
\bibliographystyle{h-physrev}
\bibliography{bolognesi_PICProceedings.bib}

\begin{thebibliography}{10}

\bibitem{Evans:2008zzb}
L.~Evans and P.~Bryant,
\newblock JINST {\bf 3}, S08001 (2008).

\bibitem{Aad:2008zzm}
ATLAS Collaboration, G.~Aad {\em et~al.},
\newblock JINST {\bf 3}, S08003 (2008).

\bibitem{CMSDetector}
CMS Collaboration, R.~Adolphi {\em et~al.},
\newblock JINST {\bf 3}, S08004 (2008).

\bibitem{Wilson:1977nk}
R.~R. Wilson,
\newblock Phys.Today {\bf 30N10}, 23 (1977).

\bibitem{Blair:1996kx}
CDF-II Collaboration, R.~Blair {\em et~al.},
\newblock (1996),
\newblock {The CDF-II detector: Technical design report}.

\bibitem{Abazov:2005pn}
D0 Collaboration, V.~Abazov {\em et~al.},
\newblock Nucl.Instrum.Meth. {\bf A565}, 463 (2006), physics/0507191.

\bibitem{Melnitchouk:2011tq}
ATLAS Collaborations, CDF Collaborations, CMS Collaborations, D0
  Collaborations, LHCb Collaborations, A.~Melnitchouk, CDF, CMS, and D0,
\newblock (2011), 1111.2840,
\newblock * Temporary entry *.

\bibitem{Baak:2011ze}
M.~Baak {\em et~al.},
\newblock (2011), hep-ph/1107.0975.

\bibitem{Chlebana:2005sd}
F.~Chlebana,
\newblock AIP Conf.Proc. {\bf 792}, 380 (2005).

\bibitem{Watt:2011kp}
G.~Watt,
\newblock JHEP {\bf 1109}, 069 (2011), hep-ph/1106.5788.

\bibitem{WZCMS:2011nx}
CMS Collaboration, S.~Chatrchyan {\em et~al.},
\newblock JHEP {\bf 1110}, 132 (2011), 1107.4789.

\bibitem{Aad:2011dm}
ATLAS Collaboration, G.~Aad {\em et~al.},
\newblock (2011), hep-ex/1109.5141.

\bibitem{Chatrchyan:2011wt}
CMS Collaboration,
\newblock (2011), hep-ex/1110.4973.

\bibitem{Han:2011vw}
CDF Collaboration, J.~Han,
\newblock (2011), hep-ex/1110.0153.

\bibitem{Placakyte:2011az}
R.~Placakyte,
\newblock (2011), hep-ph/1111.5452.

\bibitem{Aad:2011gj}
ATLAS Collaboration, G.~Aad {\em et~al.},
\newblock (2011), hep-ex/1107.2381.

\bibitem{Frixione:2002ik}
S.~Frixione and B.~R. Webber,
\newblock JHEP {\bf 0206}, 029 (2002), hep-ph/0204244.

\bibitem{Frixione:2007vw}
S.~Frixione, P.~Nason, and C.~Oleari,
\newblock JHEP {\bf 0711}, 070 (2007), hep-ph/0709.2092.

\bibitem{Balazs:1997xd}
C.~Balazs and C.~Yuan,
\newblock Phys.Rev. {\bf D56}, 5558 (1997), hep-ph/9704258.

\bibitem{Mangano:2002ea}
M.~L. Mangano, M.~Moretti, F.~Piccinini, R.~Pittau, and A.~D. Polosa,
\newblock JHEP {\bf 0307}, 001 (2003), hep-ph/0206293.

\bibitem{Alwall:2007st}
J.~Alwall {\em et~al.},
\newblock JHEP {\bf 0709}, 028 (2007), hep-ph/0706.2334.

\bibitem{Gleisberg:2008ta}
T.~Gleisberg {\em et~al.},
\newblock JHEP {\bf 0902}, 007 (2009), hep-ph/0811.4622.

\bibitem{Abazov:2010mk}
D0 Collaboration, V.~M. Abazov {\em et~al.},
\newblock Phys.Rev.Lett. {\bf 106}, 122001 (2011), hep-ex/1010.0262.

\bibitem{Wjets:2010pg}
ATLAS Collaboration, G.~Aad {\em et~al.},
\newblock Phys.Lett. {\bf B698}, 325 (2011), hep-ex/1012.5382.

\bibitem{Abazov:2011rf}
D0 Collaboration, V.~M. Abazov {\em et~al.},
\newblock Phys.Lett.B  (2011), hep-ex/1106.1457.

\bibitem{ZjetsCDF}
CDF collaboration,
\newblock (2010),
\newblock CDF Conference Note 10216.

\bibitem{Abazov:2009pp}
D0 Collaboration, V.~M. Abazov {\em et~al.},
\newblock Phys.Lett. {\bf B682}, 370 (2010), hep-ex/0907.4286.

\bibitem{Chatrchyan:2011ne}
CMS Collaboration,
\newblock (2011), hep-ex/1110.3226.

\bibitem{Aad:2011jn}
ATLAS Collaboration, G.~Aad {\em et~al.},
\newblock (2011), hep-ex/1109.1403.

\bibitem{ZbjetsCMS}
CMS collaboration,
\newblock (2010),
\newblock CMS PAS EWK-10-015.

\bibitem{Aad:2011kp}
ATLAS Collaboration, G.~Aad {\em et~al.},
\newblock (2011), hep-ex/1109.1470.

\bibitem{WcjetsCMS}
CMS collaboration,
\newblock (2011),
\newblock CMS PAS EWK-11-013.

\bibitem{Abazov:2010ix}
D0 Collaboration, V.~M. Abazov {\em et~al.},
\newblock Phys.Rev. {\bf D83}, 031105 (2011), hep-ex/1010.6203.

\bibitem{Abazov:2008qz}
D0 Collaboration, V.~Abazov {\em et~al.},
\newblock Phys.Lett. {\bf B666}, 23 (2008), hep-ex/0803.2259.

\bibitem{ZbjetsCDF}
CDF collaboration,
\newblock (2011),
\newblock CDF Conference Note 10594.

\bibitem{WcjetsCDF}
CDF collaboration,
\newblock (2011),
\newblock CDF Public Note.

\bibitem{Chatrchyan:2011rr}
CMS Collaboration, S.~Chatrchyan {\em et~al.},
\newblock Phys. Lett. B701, {\bf 535-555} (2011), hep-ex/1105.2758.

\bibitem{Aad:2011tc}
ATLAS Collaboration, G.~Aad {\em et~al.},
\newblock (2011), hep-ex/1106.1592.

\bibitem{Abazov:2011rk}
The D0 Collaboration, V.~M. Abazov {\em et~al.},
\newblock (2011), hep-ex/1109.4432.

\bibitem{Aaltonen:2011zc}
CDF Collaboration, T.~Aaltonen {\em et~al.},
\newblock Phys.Rev.Lett. {\bf 107}, 051802 (2011), hep-ex/1103.2990.

\bibitem{Yang:2011db}
ATLAS Collaboration, H.~Yang,
\newblock (2011), hep-ex/1109.2576.

\bibitem{VVCMS}
CMS collaboration,
\newblock (2011),
\newblock CMS PAS EWK-11-010.

\bibitem{WZATLAS}
ATLAS collaboration,
\newblock (2011),
\newblock ATLAS-CONF-2011-099.

\bibitem{Abazov:2010qn}
D0 Collaboration, V.~M. Abazov {\em et~al.},
\newblock Phys.Lett. {\bf B695}, 67 (2011), hep-ex/1006.0761.

\bibitem{WZleptCDF}
CDF collaboration,
\newblock (2011),
\newblock CDF Note 10176.

\bibitem{Aaltonen:2009vh}
CDF Collaboration, T.~Aaltonen {\em et~al.},
\newblock Phys.Rev.Lett. {\bf 104}, 101801 (2010), hep-ex/0911.4449.

\bibitem{Abazov:2009jf}
D0 Collaboration, V.~Abazov {\em et~al.},
\newblock Phys.Rev.Lett. {\bf 104}, 071801 (2010), hep-ex/0912.5285.

\bibitem{Aaltonen:2011uj}
CDF Collaboration, T.~Aaltonen {\em et~al.},
\newblock Phys.Rev.Lett.  (2011), hep-ex/1108.2060.

\bibitem{Aad:2011xj}
ATLAS Collaboration, G.~Aad {\em et~al.},
\newblock (2011), hep-ex/1110.5016.

\bibitem{Abazov:2011td}
D0 Collaboration, V.~M. Abazov {\em et~al.},
\newblock Phys.Rev. {\bf D84}, 011103 (2011), hep-ex/1104.3078.

\bibitem{WZCDF}
CDF collaboration,
\newblock (2011),
\newblock CDF/PUB/EXOTICS/PUBLIC/10603.

\end{thebibliography}
%\begin{thebibliography}{9}   % Use for  1-9  references
%\bibitem{tom} T.~Junk, Nucl. Instrum. Meth. A {\bf 434}, 435 (1999)
%\end{thebibliography}

\end{document}